%% file: CIKM2021.tex
\PassOptionsToPackage{table,xcdraw}{xcolor}

\documentclass[sigconf,authorversion]{acmart}

\usepackage[table,xcdraw]{xcolor}
\usepackage{multicol}
\usepackage{multirow}
\usepackage[normalem]{ulem}
\useunder{\uline}{\ul}{}
\usepackage{hyperref}
\usepackage{graphicx}
\usepackage{subfigure}
\usepackage{fixltx2e}

\copyrightyear{2021}
\acmYear{2021}
\setcopyright{acmcopyright}\acmConference[CIKM '21]{Proceedings of the 30th ACM
International Conference on Information and Knowledge Management}{November
1--5, 2021}{Virtual Event, QLD, Australia}
\acmBooktitle{Proceedings of the 30th ACM International Conference on Information
and Knowledge Management (CIKM '21), November 1--5, 2021, Virtual Event, QLD,
Australia}
\acmPrice{15.00}
\acmDOI{10.1145/3459637.3482291}
\acmISBN{978-1-4503-8446-9/21/11}

\begin{document}
\fancyhead{}
\title{UltraGCN: Ultra Simplification of Graph Convolutional Networks for Recommendation}

\author{Kelong Mao}
\authornote{~Part of the work was done during the internship at Huawei Noah's Ark Lab when the author studied at Tsinghua University.}
\affiliation{Gaoling School of AI}
\affiliation{Renmin University of China}
\affiliation{\texttt{mkl@ruc.edu.cn}}

\author{Jieming Zhu}
\affiliation{Huawei Noah's Ark Lab}
\affiliation{Shenzhen, China}
\affiliation{\texttt{jiemingzhu@ieee.org}}

\author{Xi Xiao}
\authornote{~Corresponding Author}
\affiliation{Tsinghua University}
\affiliation{Peng Cheng Laboratory}
\affiliation{\texttt{xiaox@sz.tsinghua.edu.cn}}

\author{Biao Lu}
\affiliation{Huawei Noah's Ark Lab, China}
\affiliation{\texttt{lubiao4@huawei.com}}

\author{Zhaowei Wang}
\affiliation{Huawei Noah's Ark Lab}
\affiliation{\texttt{wangzhaowei3@huawei.com}}

\author{Xiuqiang He}
\affiliation{Huawei Noah's Ark Lab}
\affiliation{\texttt{hexiuqiang1@huawei.com}\vspace{2ex}}







\begin{abstract}
    With the recent success of graph convolutional networks (GCNs), they have been widely applied for recommendation, and achieved impressive performance gains. The core of GCNs lies in its message passing mechanism to aggregate neighborhood information. However, we observed that message passing largely slows down the convergence of GCNs during training, especially for large-scale recommender systems, which hinders their wide adoption. LightGCN makes an early attempt to simplify GCNs for collaborative filtering by omitting feature transformations and nonlinear activations. In this paper, we take one step further to propose an ultra-simplified formulation of GCNs (dubbed UltraGCN), which skips infinite layers of message passing for efficient recommendation.  
    Instead of explicit message passing, UltraGCN resorts to directly approximate the limit of infinite-layer graph convolutions via a constraint loss. Meanwhile, UltraGCN allows for more appropriate edge weight assignments and flexible adjustment of the relative importances among different types of relationships. This finally yields a simple yet effective UltraGCN model, which is easy to implement and efficient to train. Experimental results on four benchmark datasets show that UltraGCN not only outperforms the state-of-the-art GCN models but also achieves more than 10x speedup over LightGCN. Our source code will be available at \textcolor{magenta}{\url{https://reczoo.github.io/UltraGCN}}.
\end{abstract}


%



\maketitle

\input{sections/1_introduction.tex}

\input{sections/2_background.tex}

\input{sections/3_approach.tex}
\input{sections/4_experiment.tex}

\input{sections/5_relatedwork.tex}

\input{sections/6_conclusion.tex}
\input{sections/7_appendix.tex}

\balance

\bibliographystyle{ACM-Reference-Format}
\bibliography{CIKM2021}
\end{document}

%% file: sections/1_introduction.tex
\section{Introduction}\label{introduction}

Nowadays, personalized recommendation has become a prevalent way to help users find information of their interests in various applications, such as e-commerce, online news, and social media. The core of recommendation is to precisely match a user's preference with candidate items. Collaborative filtering (CF)~\cite{NeuMF}, as a fundamental recommendation task, has been widely studied in both academia and industry. A common paradigm of CF is to learn vector representations (i.e., embeddings) of users and items from historical interaction data and then perform top-k recommendation based on the pairwise similarity between user and item embeddings.


As the interaction data can be naturally modelled as graphs, such as user-item bipartite graph and item-item co-occurrence graph, recent studies~\cite{NGCF,LightGCN,NIA-GCN,DHCF} opt for powerful graph convolutional/neural networks (GCNs, or GNNs in general) to learn user and item node representations. These GCN-based models are capable of exploiting higher-order connectivity between users and items, and therefore have achieved impressive performance gains for recommendation. PinSage~\cite{PinSage} and M2GRL~\cite{M2GRL} are two successful use cases in industrial applications. 

Despite the promising results obtained, we argue that current model designs are heavy and burdensome. In order to capture higher-order collaborative signals and better model the interaction process between users and items, current GNN-based CF models~\cite{GC-MC, NGCF, NIA-GCN, LCFN} tend to seek for more and more sophisticated network encoders. However, we observed that these GCN-based models are hard to train with large graphs, which hinders their wide adoption in industry. Industrial recommender systems usually involve massive graphs due to the large numbers of users and items. This brings efficiency and scalability challenges for model designs. Towards this end, some research efforts~\cite{LR-GCCF, LightGCN, RGCF} have been made to simplify the design of GCN-based CF models, mainly by removing feature transformations and non-linear activations that are not necessary for CF. These simplified models not only obtain much better performance than those complex ones, but also brings some benefits on training efficiency. 

Inspired by these pioneer studies, we performed further empirical analysis on the training process of GCN-based models and found that message passing (i.e., neighborhood aggregation) on a large graph is usually time-consuming for CF. In particular, stacking multiple layers of message passing could lead to the slow convergence of GCN-based models on CF tasks. Although the aforementioned models such as LightGCN~\cite{LightGCN} have already been simplified for training, the message passing operations still dominate their training. For example, in our experiments, three-layer LightGCN takes more than 700 epochs to converge to its best result on the Amazon-Books dataset~\cite{amazonbooks}, which would be unacceptable in an industrial setting. How to improve the efficiency of GCN models yet retain their effectiveness on recommendation is still an open problem.

%

To tackle this challenge, in this work, we question the necessity of explicit message passing layers in CF, and finally propose an ultra-simplified form of GCNs (dubbed UltraGCN) without message passing for efficient recommendation. More specifically, we analyzed the message passing formula of LightGCN and identified three critical limitations: 1) The weights assigned on edges during message passing are counter-intuitive, which may not be appropriate for CF. 2) The propagation process recursively combines different types of relationship pairs (including user-item pairs, item-item pairs, and user-user pairs) into the model, but fails to capture their varying importance. This may also introduce noisy and uninformative relationships that  confuse the model training. 
3) The over-smoothing issue limits the use of too many layers of message passing in LightGCN. Therefore, instead of performing explicit message passing, we seek to directly approximate the limit of infinite-layer graph convolutions via a constraint loss, which leads to the ultra-simplified GCN model, UltraGCN. The loss-based design of UltraGCN is very flexible, allowing us to manually adjust the relative importances of different types of relationships and also avoid the over-smoothing problem by negative sampling. This finally yields a simple yet effective UltraGCN model, which is easy to implement and efficient to train. Furthermore, we show that UltraGCN achieves significant improvements over the state-of-the-art CF models. For instance, UltraGCN attains up to 76.6\% improvement in NDCG@20 and more than 10x speedup in training over LightGCN on the Amazon-Books dataset.


In summary, this work makes the following main contributions:
\begin{itemize}
	\item We empirically analyze the training inefficiency of LightGCN and further attribute its cause to the critical limitations of the message passing mechanism. 
    \item We propose an ultra simplified formulation of GCN, namely UltraGCN, which skips infinite layers of explicit message passing for efficient recommendation.
    \item Extensive experiments have been conducted on four benchmark datasets to show the effectiveness and efficiency of UltraGCN. 
\end{itemize}


%% file: sections/2_background.tex
\section{Motivation} \label{motivation}
In this section, we revisit the GCN and LightGCN models, and further identify the limitations resulted from the inherent message passing mechanism, which also justify the motivation of our work.

\subsection{Revisiting GCN and LightGCN}
GCN~\cite{GCN} is a representative model of graph neural networks that applies message passing to aggregate neighborhood information. The message passing layer with self-loops is defined as follows:
\begin{equation}
E^{(l+1)} = \sigma \Big(\hat{D}^{-\frac{1}{2}}\hat{A}\hat{D}^{-\frac{1}{2}}E^{(l)}W^{(l)} \Big)
\end{equation}
with $\hat{A} = A + I$ and $\hat{D} = D + I$. $A$, $D$, $I$ are the adjacency matrix, the diagonal node degree matrix, and the identity matrix, respectively. $I$ is used to integrate self-loop connections on nodes. $E^{(l)}$ and $W^{(l)}$ denote the representation matrix and the weight matrix for the $l$-th layer. $\sigma(\cdot)$ is a non-linear activation function (e.g., ReLU). 

Despite the wide success of GCN in graph learning, several recent studies~\cite{LightGCN,LR-GCCF,RGCF,SGCN} found that simplifying GCN appropriately can further boost the performance on CF tasks. LightGCN~\cite{LightGCN} is one such simplified GCN model that removes feature transformations (i.e., $W^{(l)}$) and non-linear activations (i.e., $\sigma$). Its message passing layer can thus be expressed as follows:
\begin{equation}
E^{(l+1)} = ({\hat{D}}^{-\frac{1}{2}}\hat{A}{\hat{D}}^{-\frac{1}{2}})E^{(l)}\label{LightGCN}
\end{equation}
It is worth noting that although LightGCN also removes self-loop connections on nodes, its layer combination operation has a similar effect to self-loops used in Equation 2, becauase both of them output a weighted sum of the embeddings propagated at each layer as the final output representation. Given self-loop connections, we can rewrite the message passing operations for user $u$ and item $i$ as follows:
\begin{eqnarray}
e_u^{(l+1)} &=& \frac{1}{{d}_u + 1} e_{u}^{(l)} + \sum_{k \in \mathcal{N}(u)} \frac{1}{\sqrt{{d}_u + 1}\sqrt{ {d}_k + 1}} e_k^{(l)}, \label{message_passing}\\
e_i^{(l+1)} &=& \frac{1}{{d}_i + 1} e_{i}^{(l)} + \sum_{v \in \mathcal{N}(i)} \frac{1}{\sqrt{{d}_v + 1}\sqrt{ {d}_i + 1}} e_v^{(l)} \label{message_passing2}
\end{eqnarray}
where $e_{u}^{(l)}$ and $e_{u}^{(l)}$ denote the embeddings of user $u$ and item $i$ at layer $l$. $\mathcal{N}(u)$ and $\mathcal{N}(i)$ represent their neighbor node sets, respectively. ${d}_u$ denotes the original degree of the node $u$. 

As shown in the left part of Figure~\ref{architecture}, LightGCN performs a stack of message passing layers to obtain the embeddings and finally uses their dot product for training.

\subsection{Limitations of Message Passing}
\label{limitation_of_mp}
We argue that such message passing layers have potential limitations that hinder the effective and efficient training of GCN-based models in recommendation tasks. To illustrate it, we take the $l$-th layer message passing of LightGCN in Equation~\ref{message_passing} and \ref{message_passing2} for example. Note that $u$ and $v$ denote users while $i$ and $k$ denote items. LightGCN takes the dot product of the two embedding as the final logit to capture the preference of user $u$ on item $i$. Thus we obtain:
\begin{eqnarray}
\begin{aligned}
e_u^{(l+1)} \cdot e_i^{(l+1)} =  \alpha_{ui} (e_u^{(l)} \cdot e_i^{(l)}) ~~+ \sum_{k \in \mathcal{N}(u)}\alpha_{ik}(e_i^{(l)} \cdot e_k^{(l)} )~~~~ + \\ 
\sum_{v \in \mathcal{N}(i)}\alpha_{uv}(e_u^{(l)} \cdot e_v^{(l)}) ~~+
 \sum_{k \in \mathcal{N}(u)}\sum_{v \in \mathcal{N}(i)} \alpha_{kv} (e_k^{(l)} \cdot e_v^{(l)})~~,\label{one_layer_dot_production}	
\end{aligned}
\end{eqnarray}
where $\alpha_{ui}$, $\alpha_{ik}$, $\alpha_{uv}$, and $\alpha_{kv}$ can be derived as follows:
\begin{equation}
	\begin{aligned}
			\alpha_{ui} &= \frac{1}{(d_u + 1)(d_i + 1)}~, \\ 
	\alpha_{ik} &= \frac{1}{\sqrt{{d}_u + 1}\sqrt{ {d}_k + 1}(d_i + 1)}~, \\ 
	\alpha_{uv} &= \frac{1}{\sqrt{{d}_v + 1}\sqrt{ {d}_i + 1}(d_u + 1)}~, \\ 
	\alpha_{kv} &= \frac{1}{\sqrt{{d}_u + 1}\sqrt{ {d}_k + 1}\sqrt{{d}_v + 1}\sqrt{ {d}_i + 1}}  \nonumber
	\end{aligned}
\end{equation}
Therefore, we can observe that multiple different types of collaborative signals, including user-item relationships ($u$-$i$ and $k$-$v$), item-item relationships ($k$-$i$), and user-user relationships ($u$-$v$), are captured when training GCN-based models with message passing layers. 
This also reveals why GCN-based models are effective for CF. 

However, we found that the edge weights assigned on various types of relationships are not justified to be appropriate for CF tasks. Based on our empirical analysis, we identify three critical limitations of the message passing layers in GCN-based models: 
\begin{itemize}
	\item \textbf{Limitation I}: The weight $\alpha_{ik}$ is used to model the item-item relationships. However, given the user $u$, the factors of item $i$ and item $k$ are asymmetric ($\frac{1}{\sqrt{d_k + 1}}$ for item $k$ while $\frac{1}{d_i + 1}$ for item $i$). This is not reasonable since it is counter-intuitive to treat the item $k$ and item $i$ unequally. 
	Similarly, $\alpha_{uv}$ that models the user-user relationships also suffer this issue. Such unreasonable weight assignments may mislead the model training and finally result in sub-optimal performance.
	\item \textbf{Limitation II}: The message passing recursively combine different types of relationships into the modeling. While such collaborative signals should be beneficial, the above message passing formula fails to capture their varying importance. Meanwhile, stacking multiple layers of message passing as in Equation~\ref{one_layer_dot_production} likely introduce uninformative, noisy, or ambiguous relationships, which could largely affect the training efficiency and effectiveness. It is desirable to flexibly adjust the relative importances of various relationships. We validate this empirically in Section~\ref{sec::ablation_study}.
    \item \textbf{Limitation III}: Stacking more layers of message passing should capture higher-order collaborative signals, but in fact the performance of LightGCN begins to degrade at layer 2 or 3~\cite{LightGCN}. We partially attribute it to the over-smoothing problem of message passing. As graph convolution is a special form of Laplacian smoothing~\cite{deeper_insights_GCN}, performing too many layers of message passing will make the nodes with the same degrees tend to have exactly the same embeddings. According to Theorem 1 in~\cite{GCNII}, we can derive the infinite powers of message passing which take the following limit:
    \begin{equation}
    	 \lim_{l \to \infty} (\hat{D}^{-\frac{1}{2}}\hat{A}\hat{D}^{-\frac{1}{2}})^{l}_{i, j} = \frac{\sqrt{(d_i + 1)(d_j + 1)}}{2m + n}
    	 \label{infinite_mp}
    \end{equation}
    where $n$ and $m$ are the total numbers of nodes and edges in the graph, respectively.



\end{itemize}


The above limitations of message passing motivate our work. We question the necessity of explicit message passing layers in CF and further propose an ultra-simplified formulation of GCN, dubbed UltraGCN.

%% file: sections/3_approach.tex
\begin{figure}[!t]
	\centering
	\includegraphics[width=0.49\textwidth]{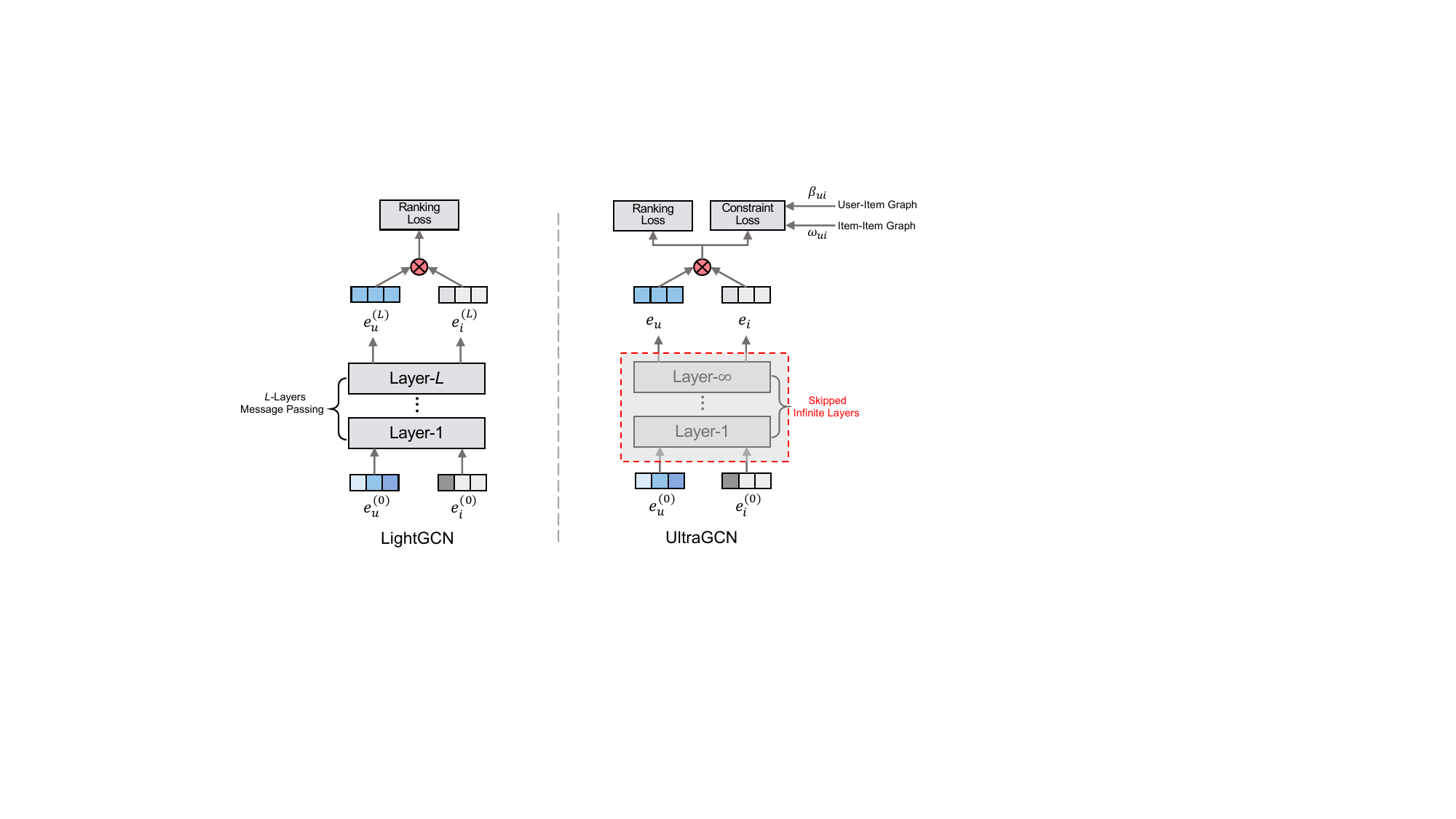}
	\caption{Illustrations of training of LightGCN (left) and UltraGCN (right). LightGCN needs to recurrently perform $L$-layers message passing to get the final embeddings for training, while UltraGCN can ``skip'' such message passing to make the embeddings be directly trained, largely improving training efficiency and helping real deployment.}\label{architecture}
\end{figure}
%
%



\section{U\lowercase{ltra}GCN}

In this section, we present our ultra-simplified UltraGCN model and 
demonstrate how to incorporate different types of relationships in a flexible manner. We also elaborate on how it overcomes the above limitations and analyze its connections to other related models.

\subsection{Learning on User-Item Graph}
\label{learning_on_ui}

Due to the limitations of message passing, in this work, we take one step forward to question the necessity of explicit message passing in CF. Considering that the limit of infinite powers of message passing exists as shown in Equation~\ref{infinite_mp}, we wonder whether it is possible to skip the infinite-layer message passing yet approximate the convergence state reached. 

After repeating infinite layers of message passing, we express the final convergence condition as follows:
\begin{eqnarray}
&&
e_u = \lim_{l \to \infty}e_u^{(l+1)} = \lim_{l \to \infty}e_u^{(l)}
\end{eqnarray}
That is, the representations of the last two layers keep unchanged, since the vector generated from neighborhood aggregation equals to the node representation itself. We use $e_u$ (or $e_i$) to denote the final converged representation of user $u$ (or item $i$). Then, Equation~\ref{message_passing} can be rewritten as:
\begin{equation}
e_u = \frac{1}{{d}_u + 1} e_{u} + \sum_{i \in \mathcal{N}(u)} \frac{1}{\sqrt{{d}_u + 1}\sqrt{ {d}_i + 1}} e_i \label{message_passing3}
\end{equation}
After some simplifications, we derive the following convergence state:
\begin{equation}
e_u = \sum_{i \in \mathcal{N}(u)} \beta_{u, i} e_i \label{termination}~, \:\:\: \beta_{u, i} = \frac{1}{d_u} \sqrt{\frac{{d}_u + 1}{{d}_i + 1}}
\end{equation}
In other words, if Equation~\ref{termination} is satisfied for each node, it reaches the convergence state of message passing. 

Instead of performing explicit message passing, we aim to directly approximate such convergence state. To this end, a straightforward way is to minimize the difference of both sides of Equation~\ref{termination}. In this work, we normalize the embeddings to unit vectors and then maximize the dot product of both terms:
\begin{eqnarray}
     \operatorname{max} \sum_{i \in \mathcal{N}(u)} \beta_{u, i} e_u^{\top} e_i~,\:\:\: \forall u \in U \:, \label{structure_target}
\end{eqnarray}
which is equivalent to maximize the cosine similarity between $e_u$ and $e_i$. For ease of optimization, we further incorporate sigmoid activation and negative log likelihood~\cite{cross_entropy_support}, and derive the following loss:
\begin{eqnarray}
\mathcal{L}_C = -\sum_{u \in U} \sum_{i \in \mathcal{N}(u)} \beta_{u, i} \log \big(\sigma(e_u^{\top} e_{i}) \big) \:, \label{final_UltraGCN}
\end{eqnarray}
where $\sigma$ is the sigmoid function. The loss is optimized to fulfill the structure constraint imposed by Equation~\ref{termination}. As such, we denote $\mathcal{L}_C$ as the \textit{constraint loss} and denote $\beta_{u,i}$ as the \textit{constraint coefficient}. 


However, optimizing $\mathcal{L}_C$ could also suffer from the over-smoothing problem as $\mathcal{L}_C$ requires all connected pairs ($\beta_{u,i} > 0$) to be similar. In this way, users and items could easily converge to the same embeddings. 
To alleviate the over-smoothing problem, conventional GCN-based CF models usually fix a small number of message passing layers, e.g., 2$\sim$4 layers in LightGCN. Instead, as UltraGCN approximates the limit of infinite-layer message passing via a constraint loss, we choose to perform negative sampling during training. This is inspired from the negative sampling strategy used in Word2Vec~\cite{word2vec}, which provides a more simple and effective way to counteract the over-smoothing problem. 
After performing negative sampling, we finally derive the following constraint loss:
\begin{eqnarray}
\begin{split}
\mathcal{L}_C=-\hspace{-2ex}\sum_{(u,i)\in N^+}\hspace{-2ex}\beta_{u,i} \log \big(\sigma(e_u^{\top} e_{i})\big) -\hspace{-2ex}\sum_{(u,j)\in N^-}  \hspace{-2ex}\beta_{u,j} \log \big(\sigma(-e_u^{\top} e_{j} )\big)\label{Lc_loss}
\end{split}
\end{eqnarray}
where $N^+$ and $N^-$ represent the sets of positive pairs and randomly sampled negative pairs. Note that we omit the summation over $U$ for ease of presentation.
The constraint loss $\mathcal{L}_C$ enables UltraGCN to directly approximate the limit of infinite-layer message passing to capture arbitrary high-order collaborative signals in the user-item bipartite graph, while effectively avoiding the troublesome over-smoothing issue via negative sampling.  
Furthermore, we note that $\beta_{u, i}$ acts as the loss weight in $\mathcal{L}_C$, which is inversely proportional to ${d}_u$ and ${d}_i$ with similar magnitudes. This is interpretable for CF. If a user interacts with many items or an item is interacted by many users, the influence of their interaction would be small, and thus the loss weight of this ($u$, $i$) pair should be small.

\subsubsection{Optimization}
Typically, CF models perform item recommendation by applying either pairwise BPR (Bayesian personalized ranking) loss~\cite{BPR} or pointwise BCE (binary cross-entropy) loss~\cite{NeuMF} for optimization. We formulate CF as a link prediction problem in graph learning. Therefore, we choose the following BCE loss as the main optimization objective. It is also consistent with the loss format of $\mathcal{L}_C$.
\begin{eqnarray}
\mathcal{L}_{O} =  -\hspace{-2ex}\sum_{(u,i)\in N^+}\hspace{-2ex}\log \big(\sigma(e_u^{\top} e_{i})\big) - \hspace{-2ex}\sum_{(u,j)\in N^-}\hspace{-2ex} \log \big(\sigma(-e_u^{\top} e_{j}  )\big)
\end{eqnarray}
where $N^+$ and $N^-$ represent positive and randomly sampled negative links (i.e., $u$-$j$ pairs). Note that for simplicity, we use the same sets of sample pairs with $\mathcal{L}_C$, but they could also be made different conveniently. 

As $\mathcal{L}_{O}$ and $\mathcal{L}_{C}$ depends only on the user-item relationships, we define it as the base version of UltraGCN, denoted as  UltraGCN$_{Base}$, which has the following optimization objective. 
\begin{eqnarray}
\mathcal{L} = \mathcal{L}_{O} + \lambda \mathcal{L}_C \:,
\end{eqnarray}
where $\lambda$ is the hyper-parameter to control the importance weights of two losse terms.

\subsection{Learning on Item-Item Graph}
\label{learning_on_ii}
As Equation~\ref{one_layer_dot_production} shows, 
except for user-item relationships, some other relationships (e.g., item-item and user-user relationships) also greatly contribute to the effectiveness of GCN-based models on CF. However, in conventional GCN-based models, these relationships are implicitly learned through the same message passing layers with user-item relationships. This not only leads to the unreasonable edge weight assignments as discussed in Section~\ref{limitation_of_mp}, but also fails to capture the relative importances of different types of relationships. In contrast, UltraGCN does not rely on explicit message passing so that we can separately learn other relationships in a more flexible way. This also enables us to manually adjust the relative importances of different relationships.

We emphasize that UltraGCN is flexible to extend to model many different relation graphs, such as user-user graphs, item-item graphs, and even knowlege graphs. In this work, we mainly demonstrate its use on the item-item co-occurrence graph, which has been shown to be useful for recommendation in~\cite{M2GRL}. We first build the item-item co-occurrence graph by linking items that have co-occurrences, which produces the following weighted adjacent matrix $G \in \mathcal{R}^{|I| \times |I|}$. 
\begin{eqnarray}
G = A^{\top}A
\end{eqnarray}
where each entry denotes the co-occurrences of two items. We follow Equation~\ref{termination} to approximate infinite-layer graph convolution on $G$ and derive the new coefficient $\omega_{i, j}$:
\begin{eqnarray}
 \omega_{i, j} = \frac{G_{i, j}}{g_i-G_{i, i}} \sqrt{\frac{g_i}{g_{j}}}~, \:\:\: g_i=\sum_kG_{i,k}
\end{eqnarray}
where $g_i$ and $g_{j}$ denote the degrees (sum by column) of item $i$ and item $j$ in $G$, respectively.

Similar to Equation~\ref{Lc_loss}, we can derive the constraint loss on the item-item graph to learn the item-item relationships in an explicit way.
However, as the adjacency matrix $G$ of the item-item graph is usually much denser compared to the sparse adjacency matrix $A$ of the user-item graph, directly minimizing the constraint loss on $G$ would introduce too many unreliable or noisy item-item pairs into optimization, which may make UltraGCN difficult to train. This is also similar to the \textbf{Limitation II} of conventional GCN-based models described in Section~\ref{limitation_of_mp}. But thanks to the flexible design of UltraGCN, we choose to select only informative pairs for training.  

Specifically, to keep sparse item connections and retain training efficiency, we first select top-$K$ most similar items $S(i)$ for item $i$ according to $\omega_{i, j}$.
Intuitively, $\omega_{i, j}$ measures the similarity of item $i$ and item $j$, since it is proportional to the co-occurrence number of item $i$ and item $j$, yet inversely proportional to the total degrees of both items. Instead of directly learning item-item pairs, we propose to augment positive user-item pairs to capture item-item relationships. This keeps the training terms of UltraGCN being unified and decrease the possible difficulty in multi-task learning. We also empirically show that such a way can achieve better performance in Section~\ref{sec::ablation_study}.   
For each positive ($u$, $i$) pair, we first  construct $K$ weighted positive ($u$, $j$) pairs, for $j\in S(i)$. Then, we penalize the learning of these pairs with the more reasonable similarity score $\omega_{i, j}$ and derive the constraint loss $\mathcal{L}_{I}$ on the item-item graph as follow:
\begin{eqnarray}
\mathcal{L}_{I} = -\sum_{(u,i)\in N^+}\sum_{j \in S(i)} \omega_{i, j} \log(\sigma(e_{u}^{\top} e_{j}))
\end{eqnarray} 
where $|S(i)| = K$. 
We omit the negative sampling here as the negative sampling in $\mathcal{L}_C$ and $\mathcal{L}_O$ has already enabled UltraGCN to counteract over-smoothing. 
With this constraint loss, we extend UltraGCN to better learn item-item relationships, and finally derive the following training objective of UltraGCN,
\begin{eqnarray}
\mathcal{L} = \mathcal{L}_{O} + \lambda \mathcal{L}_C + \gamma \mathcal{L}_I  
\end{eqnarray}
where $\lambda$ and $\gamma$ are hyper-parameters to adjust the relative importances of user-item and item-item relationships, respectively.

Figure~\ref{architecture} illustrates the simple architecture of UltraGCN in contrast to LightGCN. Similarly, in inference, we use the dot product $\hat{y}_{ui} = e_u^{\top} e_i$ between user $u$ and item $i$ as the ranking score for recommendation.

\subsection{Discussion}
\label{model_strength}
\subsubsection{Model Analysis}
We first analyze the strengths of our UltraGCN model: 1) The weights assigned on edges in UltraGCN, i.e., $\beta_{i, j}$ and $\omega_{i, j}$, are more reasonable and interpretable for CF, which are helpful to better learn user-item and item-item relationships, respectively. 
2) Without explicit message passing, UltraGCN is flexible to separately customize its learning with different types of relationships. It is also able to select valuable training pairs (as in Section~\ref{learning_on_ii}), rather than learn from all neighbor pairs indistinguishably, which may be mislead by noise. 3) Although UltraGCN is trained with different types of relationships in a multi-task learning way, its training losses (i.e., $\mathcal{L}_C$, $\mathcal{L}_I$, and $\mathcal{L}_{O}$) are actually unified, following the form of binary cross entropy. Such unification facilitates the training of UltraGCN, which converges fast. 4) The design of UltraGCN is flexible, by setting $\gamma$ to 0, it reduces to UltraGCN$_{Base}$, which only learns on the user-item graph. The performance comparison between UltraGCN and UltraGCN$_{Base}$ is provided in Table~\ref{overall_res}.

%

Note that in the current version, we do not incorporate the modeling of user-user relationships in UltraGCN. This is mainly because that users' interests are much broader than items' attributes. We found that it is harder to capture the user-user relationships from the user-user co-occurrence graph only. In Section~\ref{sec::ablation_study}, we empirically show that learning on the user-user co-occurrence graph does not bring noticeable improvements to UltraGCN. In contrast, conventional GCN-based CF models indistinguishably learn over all relationships from the user-item graph (i.e., Limitation II) likely suffer from performance degradation. The user-user relationships may be better modeled from 
a social network graph, and we leave it for future work.


\subsubsection{Relations to Other Models}
\label{sec::relations_to_other_models}
In this part, we discuss the relations between our UltraGCN and some other existing models.

\textbf{Relation to MF}. UltraGCN is formally to be a new weighted MF model with BCE loss tailored for CF. In contrast to previous MF models (e.g., NeuMF~\cite{NeuMF}), UltraGCN can more deeply mine the collaborative information using graphs, yet keep the same concise architecture and model efficiency as MF.

\textbf{Relation to Network Embedding Methods}. Qiu et al.~\cite{unify_network_embedding} have proved that many popular network embedding methods with negative sampling (e.g., DeepWalk~\cite{deepwalk}, LINE~\cite{LINE}, and Node2Vec~\cite{node2vec}) all can be unified into the MF framework. However, in contrast to these network embedding methods, the edge weights used in UltraGCN are more meaningful and reasonable for CF, and thus lead to much better performance. 
In addition, the random walk in many network embedding methods will also uncontrollably introduce uninformative relationships that affect the performance.
We empirically show the superiority of UltraGCN over three typical network embedding methods on CF in Section~\ref{performance_comparison}.

\textbf{Relation to One-Layer LightGCN}. We emphasize that UltraGCN is also different from one-layer LightGCN with BCE loss, because LightGCN applies weight combination to embeddings aggregation while our constraint coefficients are imposed on the constraint loss function, which aims to learn the essence of infinite-layer graph convolution. 
On the contrary, UltraGCN can overcome the limitations of one-layer LightGCN as described in Section~\ref{learning_on_ii}.



\subsubsection{Model Complexity}
\label{sec::complexity}
Given the embedding size $d$, $K$ similar items for each $S(i)$, $R$ as the number of negative samples for each positive pair, and $|A^{+}|$ as the number of valid non-zero entries in the user-item interaction matrix, we can derive the training time complexity of UltraGCN: $\mathcal{O}((K+R+1) * |A^{+}| * (d^2 + 1))$. We note that the time complexities to calculate $\beta$ and $\omega$ are  $\mathcal{O}(1)$, since we can pre-calculate them offline before training. As we usually limit $K$ to be small (e.g., 10 in our experiments) in practice, the time complexity of UltraGCN lies in the same level with MF, which is $\mathcal{O}((R + 1) * |A^{+}| * d^2)$. Besides, the only trainable parameters in UltraGCN are the embeddings of users and items, which is also the same with MF and LightGCN. As a result, our low-complexity UltraGCN brings great efficiency for model training and should be more practically applicable to large-scale recommender systems.

%% file: sections/4_experiment.tex
\section{Experiments}
We first compare UltraGCN with various state-of-the-art CF methods to demonstrate its effectiveness and high efficiency. We also perform detailed ablation studies to justify the rationality and effectiveness of the design choices of UltraGCN. 

\subsection{Experimental Setup}

\textbf{Datasets and Evaluation Protocol.} 
We use four publicly available datasets, including Amazon-Book, Yelp2018, Gowalla, and MovieLens-1M to conduct our experiments, as many recent GCN-based CF models~\cite{NGCF,LightGCN,DGCF, LCFN} are evaluated on these four datasets. 
 We closely follow these GCN-based CF studies and use the same data split as them. 
Table~\ref{data_statistics} shows the statistics of the used datasets.  

For the evaluation protocol, Recall@20 and NDCG@20 are chosen as the evaluation metrics as they are popular in the evaluation of GCN-based CF models. We treat all items not interacted by a user as the candidates, and report  
the average results over all users.


\begin{table}[!t]
\centering
\caption{Statistics of the datasets.}
\scalebox{1.0}{
\begin{tabular}{c|r|r|r|c}
\toprule
Dataset     & \#Users & \#Items & \#Interactions & Density \\ \midrule
Amazon-Book & 52, 643 & 91, 599 & 2, 984, 108    & 0.062\% \\
Yelp2018    & 31, 668 & 38, 048 & 1, 561, 406    & 0.130\% \\ 
Gowalla     & 29, 858 & 40, 981 & 1, 027, 370    & 0.084\% \\ 
Movielens-1M & 6,022 & 3,043 & 995, 154 & 5.431\% \\
\bottomrule
\end{tabular}}
\label{data_statistics}
\end{table}

\textbf{Baselines.} 
In total, we compare UltraGCN with various types of the state-of-the-art models, covering MF-based (MF-BPR~\cite{MF}, ENMF~\cite{ENMF}), metric learing-based (CML~\cite{CML}), network embedding methods (DeepWalk~\cite{deepwalk}, LINE~\cite{LINE}, and Node2Vec~\cite{node2vec}), and GCN-based (NGCF~\cite{NGCF}, NIA-GCN~\cite{NIA-GCN}, LR-GCCF~\cite{LR-GCCF}, LightGCN~\cite{LightGCN}, and DGCF~\cite{DGCF}).
\begin{table*}[!t]
\centering
\caption{Overall performance comparison. Improv. denotes the relative improvements over the best GNN-based baselines.}
\begin{tabular}{c|cc|cc|cc|cc}
\toprule
\multirow{2}{*}{Model} & \multicolumn{2}{c|}{Amazon-Books} & \multicolumn{2}{c|}{Yelp2018}     & \multicolumn{2}{c|}{Gowalla}      & \multicolumn{2}{c}{Movielens-1M}  \\ \cline{2-9} 
                       & Recall@20       & NDCG@20         & Recall@20       & NDCG@20         & F1@20           & NDCG@20         & Recall@20       & NDCG@20         \\ \midrule
MF-BPR                 & 0.0338          & 0.0261          & 0.0549          & 0.0445          & 0.1616          & 0.1366          & 0.2153          & 0.2175          \\
CML                    & \underline{0.0522}          & \underline{0.0428}          & 0.0622          & \underline{0.0536}          & 0.1670          & 0.1292          & 0.1730          & 0.1563          \\
ENMF                   & 0.0359          & 0.0281          & 0.0624          & 0.0515          & 0.1523          & 0.1315          & 0.2315          & 0.2069          \\ \hline
DeepWalk               & 0.0346          & 0.0264          & 0.0476          & 0.0378          & 0.1034          & 0.0740          & 0.1348          & 0.1057          \\
LINE                   & 0.0410          & 0.0318          & 0.0549          & 0.0446          & 0.1335          & 0.1056          & 0.2336          & 0.2226          \\
Node2Vec               & 0.0402          & 0.0309          & 0.0452          & 0.0360          & 0.1019          & 0.0709          & 0.1475          & 0.1186          \\ \hline
NGCF                   & 0.0344          & 0.0263          & 0.0579          & 0.0477          & 0.1570          & 0.1327          & 0.2513               & 0.2511               \\
NIA-GCN                & 0.0369          & 0.0287          & 0.0599          & 0.0491          & 0.1359          & 0.1106          & 0.2359          & 0.2242          \\
LR-GCCF                & 0.0335          & 0.0265          & 0.0561          & 0.0343          & 0.1519          & 0.1285          & 0.2231          & 0.2124          \\
LightGCN               & 0.0411          & 0.0315          & 0.0649          & 0.0530          & 0.1830          & 0.1554          & 0.2576          & 0.2427          \\
DGCF                   & \underline{0.0422}          & \underline{0.0324}          & \underline{0.0654}         & \underline{0.0534}          & \underline{0.1842}          & \underline{0.1561}          & \underline{0.2640}          & \underline{0.2504}          \\ \hline
UltraGCN$_{Base}$               & 0.0504 & 0.0393 & 0.0667 & 0.0552 & 0.1845 & 0.1566 & 0.2671 & 0.2539 \\
UltraGCN               & \textbf{0.0681} & \textbf{0.0556} & \textbf{0.0683} & \textbf{0.0561} & \textbf{0.1862} & \textbf{0.1580} & \textbf{0.2787} & \textbf{0.2642} \\ \hline \hline
Improv.                   &      61.37\%     & 71.60\%& 4.43\%          & 5.06\%          & 1.09\%          & 1.22\%          & 5.57\%          & 5.51\%   \\
$p$-value                   & 4.03e-8          & 5.64e-8& 1.61e-4          & 1.24e-4          & 7.21e-3          & 3.44e-4          & 4.19e-5          & 2.23e-5   \\

\bottomrule
\end{tabular}
\label{overall_res}
\end{table*}

\textbf{Parameter Settings.} 
Generally, we adopt Gaussian distribution with 0 mean and $10^{-4}$ standard deviation to initialize embeddings. In many cases, we adopt $L_2$ regularization with $10^{-4}$ weight and we set the learning rate to $10^{-4}$, the batch size to 1024, the negative sampling ratio $R$ to 300, and the size of the neighbor set $K$ to 10. In particular, we fix the embedding size to 64 which is identical to recent GCN-based work~\cite{NGCF,NIA-GCN,LightGCN,DGCF} to keep the same level of the number of parameters for fair comparison. We tune $\lambda$ in [0.2, 0.4, 0.6, 0.8, 1.0, 1.2, 1.4], and $\gamma$ in [0.1, 0.5, 1.0, 1.5, 2.0, 2.5, 3, 3.5]. For some baselines, we report the results from their papers to keep consistency. They are also comparable since we use the exactly same datasets and experimental settings provided by them. For other baselines, we mainly use their official open-source code and carefully tune the parameters to achieve the best performance for fair comparisons. To allow for reproduciblility, we have released the source code and benchmark settings of UltraGCN at RecZoo.


\subsection{Performance Comparison}
\label{performance_comparison}
Table~\ref{overall_res} reports the performance comparison results. We have the following observations:
\begin{itemize}


\item UltraGCN consistently yields the best performance across all four datasets. In particular, UltraGCN hugely improves over the strongest GCN-based baseline (i.e., DGCF) on Amazon-Book by 61.4\% and 71.6\% w.r.t. Recall@20 and NDCG@20 respectively.
The results of significance testing indicates that our improvements over the current strongest GCN-based baseline are statistically significant ($p$-value $\textless$ 0.05). With additional learning on the item-item graph, UltraGCN performs consistently better than its simpler variant UltraGCN$_{Base}$.
We attribute such good performance of UltraGCN to the following reasons:
1) Compared with
network embedding models and the other GCN-based models, UltraGCN can respectively filter uninformative user-item and item-item relationships in a soft way (i.e., optimize with $\beta$) and a hard way (i.e., only select $K$ most similar item pairs).
The edge weights for the learning of user-item and item-item relationships in UltraGCN are also more reasonable;
2) Compared with other baselines, UltraGCN can leverage powerful graph convolution to exploit useful and deeper collaborative information in graphs. These advantages together lead to the superiority of UltraGCN than compared state-of-the-art models.

\item Overall, network embedding models perform worse than GCN-based models, especially on Gowalla. The reason might be that the powerful graph convolution is more effective than traditional random walk or heuristic mining strategies in many network embedding methods, to capture collaborative information for recommendation.

\item Since UltraGCN is a special MF which only needs the dot product operation for embeddings, its architecture is orthogonal to some state-of-the-art models (e.g., DGCF). Therefore, similar to MF, UltraGCN can be deemed as an effective and efficient CF framework which is possible to be incorporated with other methods, such as enabling disentangled representation for users and items as DGCF, to achieve better performance. We leave such study in future work.
\end{itemize}

\subsection{Efficiency Comparison}
\label{efficiency_comparison}
As highlighted in Section~\ref{model_strength}, UltraGCN is endowed with high training efficiency for CF thanks to its concise and unified designs. We have also theoretically demonstrated that the training time complexity of UltraGCN is on the same level as MF in Section~\ref{sec::complexity}. In this section, we further empirically demonstrate the superiority of UltraGCN on training efficiency compared with other CF models, especially GCN-based models.
To be specific, we select MF-BPR, ENMF, LightGCN, and LR-GCCF as the competitors, which are relatively efficient models in their respective categories. To be more convincing, we compare their training efficiency from two views:
\begin{itemize}
	\item The total training time and epochs for achieving their best performance.
	\item Training them with the same epochs to see what performance they can achieve.
\end{itemize}
Note that the validation time is not included in the training time. Considering the fact that the official implementations of the compared models can be optimized to be more efficient, we use a uniform code framework implemented by ourselves for all models for fair comparison. 
In particular, our implementations refer to their official versions and optimize them with uniform acceleration methods (e.g, parallel sampling) to ensure the fairness of comparison. We will release all of our code.
Experiments are conducted on Amazon-Book with the same Intel(R) Xeon(R) Silver 4210 CPU @ 2.20GHz machine with one GeForce RTX 2080 GPU for all compared models. Results of the two experiments are shown in Table~\ref{exp_efficiency_1} and Table~\ref{exp_efficiency_2} respectively. We have the following conclusions:

(1) Table~\ref{exp_efficiency_1} shows that the training speed (i.e., Time/Epoch) of UltraGCN is close to MF-BPR, which empirically justifies our analysis that the time complexities of UltraGCN and MF are on the same level. UltraGCN needs 75 epochs to converge which is much less than LR-GCCF and LightGCN, leading to only 45 minutes for total training. Finally, UltraGCN has around 14x, 4x, 4x speedup compared with LightGCN, LR-GCCF, and ENMF respectively, demonstrating the big efficiency superiority of UltraGCN.

(2) Table~\ref{exp_efficiency_2} shows that when UltraGCN converges (i.e., train the fixed 75 epochs), the performances of all the other compared models are much worse than UltraGCN. That is to say, UltraGCN can achieve much better performance with less time, which further demonstrates the higher efficiency of UltraGCN than the other GCN-based CF models.  

\begin{table}[!t]
\centering
\caption{Efficiency comparison from the first view.}
\begin{tabular}{c|c|c|c}
\toprule
Model    & Time/Epoch & \#Epoch & Training Time \\ \midrule
MF-BPR   & \textbf{30s}        & \textbf{23}      & \textbf{12m}           \\
ENMF     & 129s       & 81     & 2h54m          \\
LR-GCCF  & 67s        & 165     & 3h5m          \\
LightGCN & 51s        & 780     & 11h3m        \\
\hline
UltraGCN & \textbf{36s}        & \textbf{75}      & \textbf{	45m}           \\ \bottomrule
\end{tabular}
\label{exp_efficiency_1}
\end{table}

\begin{table}[!t]
\caption{Efficiency comparison from the second view. All models are trained with the fixed 75 epochs except MF-BPR. Since MF-BPR needs less than 75 epochs to converge, we report its actual training time.}
\begin{tabular}{c|c|c|c}
\toprule
Model    & Training Time & Recall@20       & NDCG@20         \\ \midrule
MF-BPR   & \textbf{12m}  & 0.0338          & 0.0261          \\
ENMF     & 2h41m         & 0.0355          & 0.0277          \\
LR-GCCF  & 1h13m         &  0.0304         & 0.0185                 \\
LightGCN & 1h4m          & 0.0342          & 0.0262          \\ \hline
UltraGCN & \textbf{45m}  & \textbf{0.0681} & \textbf{0.0556} \\ \bottomrule
\end{tabular}
\label{exp_efficiency_2}
\end{table}

\subsection{Ablation Study}
\label{sec::ablation_study}
We perform ablation studies on Amazon-Book to justify the following opinions:
(i) The designs of UltraGCN is effective, which can flexibly and separately learn the user-item relationships and item-item relationships to improve recommendation performance;
(ii) Augmenting positive user-item pairs for training to learn item-item relationships can achieve better performance than optimizing between item-item pairs; 
(iii) User-user co-occurrence information is probably not very informative to help recommendation.

\begin{figure}[!t]
   \vspace{-2ex}
	\centering  
	\subfigure[Recall@20]{
		\includegraphics[width=0.485\columnwidth]{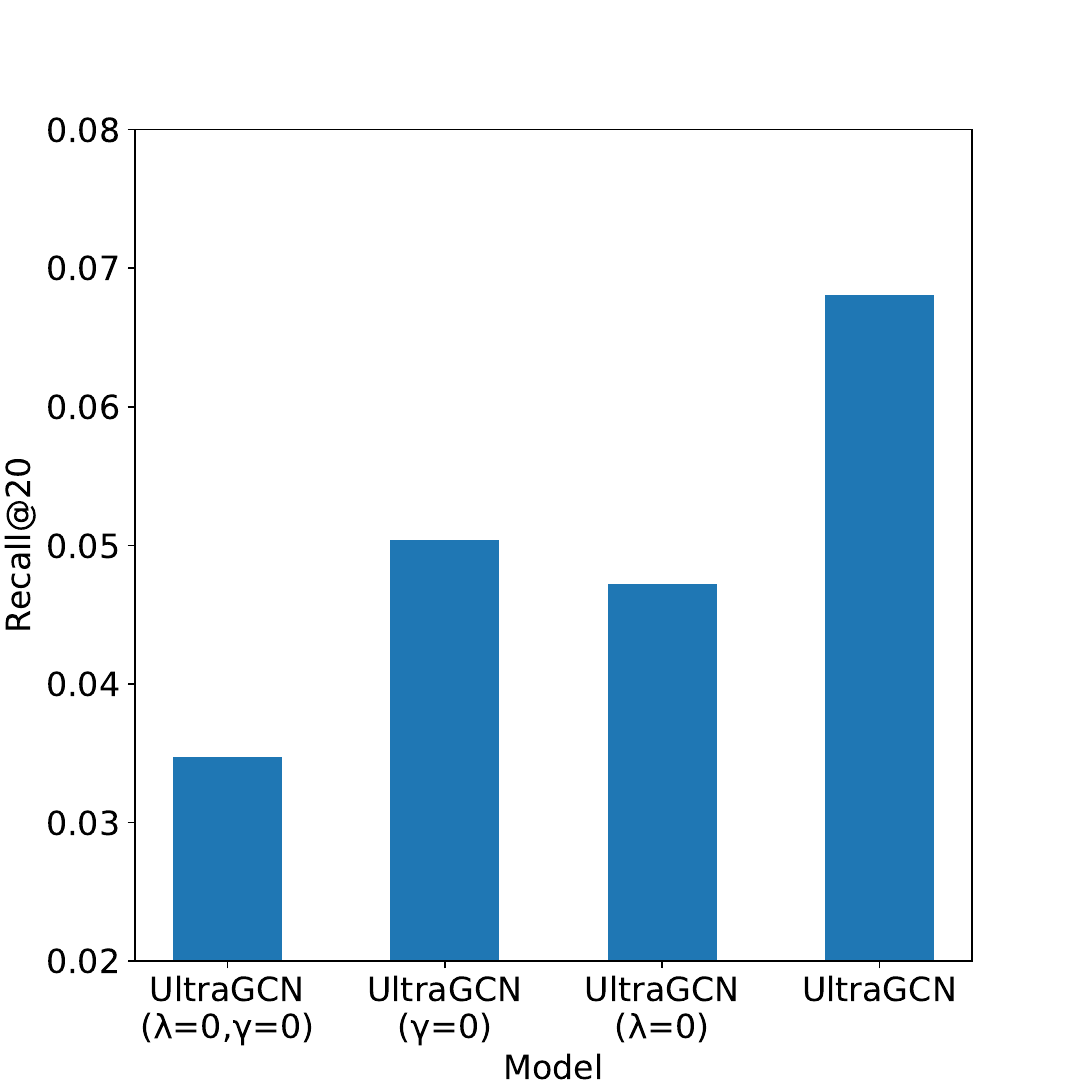}}
	\subfigure[NDCG@20]{
		\includegraphics[width=0.485\columnwidth]{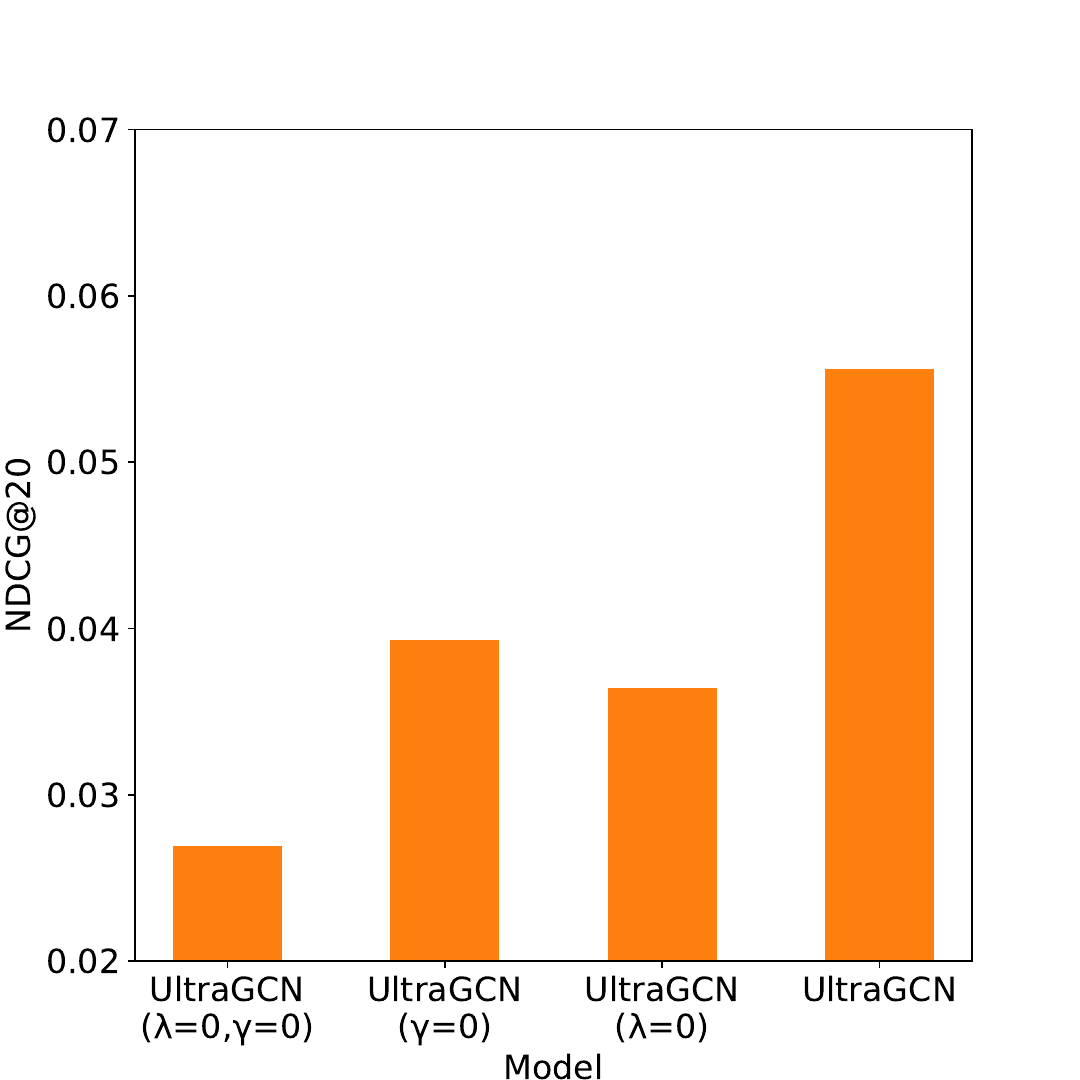}}
	\vspace{-2ex}
	\caption{Performance comparison of variants of UltraGCN.}
	\label{fig_ablation_study}
\end{figure}

\textbf{For opinion (i)}, we compare UltraGCN with the following variants to show the effectiveness of our designs in UltraGCN:
\begin{itemize}
	\item UltraGCN($\lambda=0$, $\gamma=0$): when setting $\lambda$ and $\gamma$ to 0, UltraGCN is simply reduced to MF training with BCE loss function, which does not leverage graph information and cannot capture higher-order collaborative signals.
	\item UltraGCN($\gamma=0$): this variant is identical to UltraGCN$_{Base}$, which only learns on the user-item graph and lacks more effective learning for item-item relationships.
	\item UltraGCN($\lambda=0$): this variant lacks the graph convolution ability for learning on the user-item graph to more deeply mine the collaborative information.
\end{itemize}
Results are shown in Figure~\ref{fig_ablation_study}. We have the following observations:

(1) UltraGCN($\gamma=0$) and UltraGCN($\lambda=0$) all perform better than UltraGCN($\lambda=0$, $\gamma=0$), demonstrating that the designs of UltraGCN can effectively learn on both the user-item graph and item-item graph to improve recommendation;
(2) Relatively, UltraGCN($\lambda=0$) is inferior to UltraGCN($\gamma=0$), indicating that  
user-item relationships may be better modeled than item-item relationships in UltraGCN;
(3) UltraGCN performs much better than all the other three variants, demonstrating that our idea to disassemble various relationships, eliminate uninformative things which may disturb the model learning, and finally conduct multi-task learning in a clearer way, is effective to overcome the limitations (see Section~\ref{limitation_of_mp}) of previous GCN-based CF models.

\begin{table}[!t]
\centering
\caption{Performance comparison of whether learning on the user-user co-occurrence graph.}
\scalebox{0.71}{\begin{tabular}{c|cc|cc|cc}
\toprule
\multirow{2}{*}{Model} & \multicolumn{2}{c|}{UltraGCN($\gamma=0$)} & \multicolumn{2}{c|}{UltraGCN($\lambda=0$)} & \multicolumn{2}{c}{UltraGCN} \\ \cline{2-7} 
                       & Recall@20       & NDCG@20       & Recall@20       & NDCG@20       & Recall@20      & NDCG@20     \\ \midrule
w/o $\mathcal{L}_U$   & 0.0504          & 0.0393        & 0.0472          & 0.0364        & 0.0681         & 0.0556      \\
with $\mathcal{L}_U$    & 0.0513          & 0.0399        & 0.0470          & 0.0364        & 0.0683         & 0.0559      \\ \bottomrule
\end{tabular}}
\label{table::ablation_study_uu}
\end{table}

\begin{figure}[!t]
   \vspace{-3ex}
	\centering  
	\subfigure[Recall@20]{
		\includegraphics[width=0.485\columnwidth]{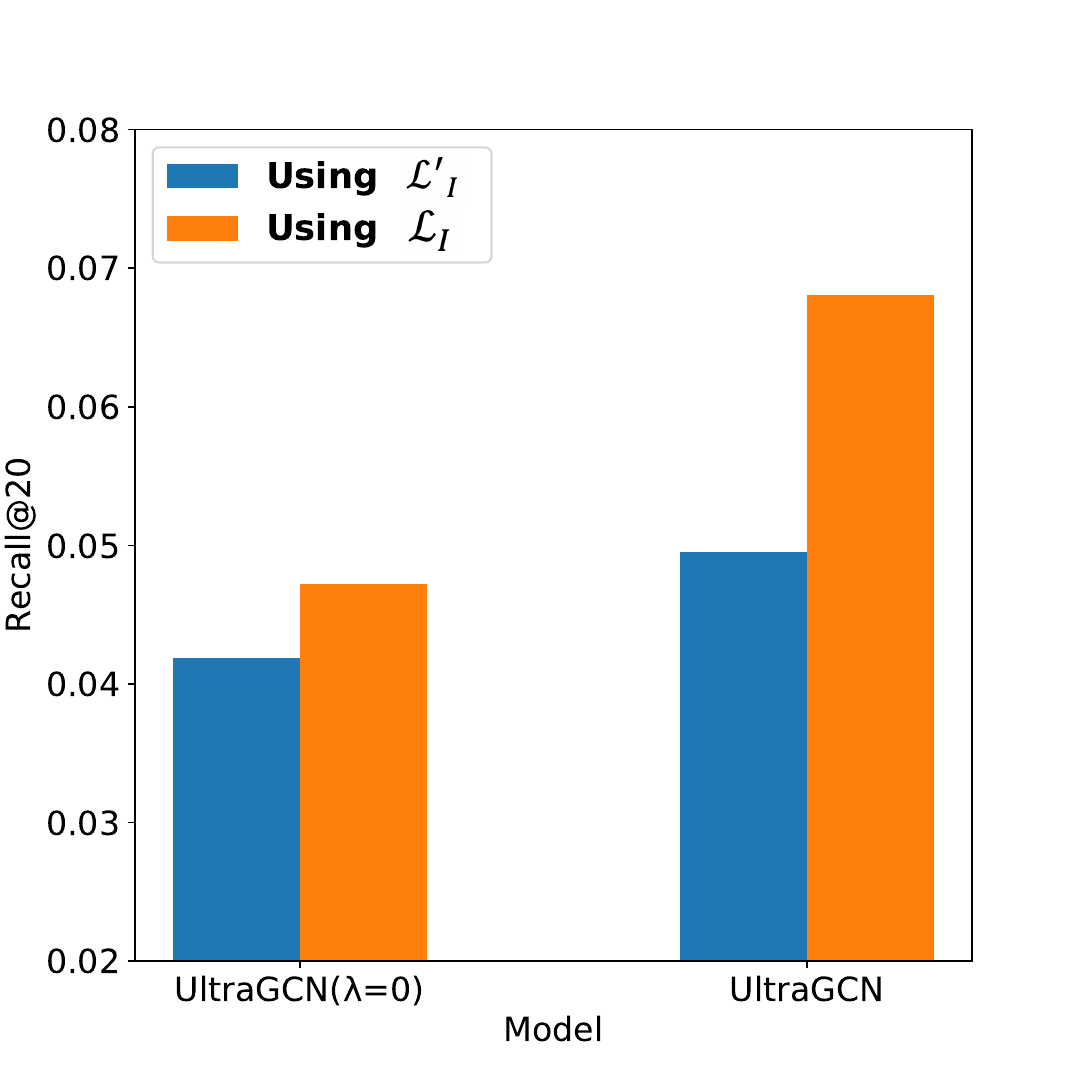}}
	\subfigure[NDCG@20]{
		\includegraphics[width=0.485\columnwidth]{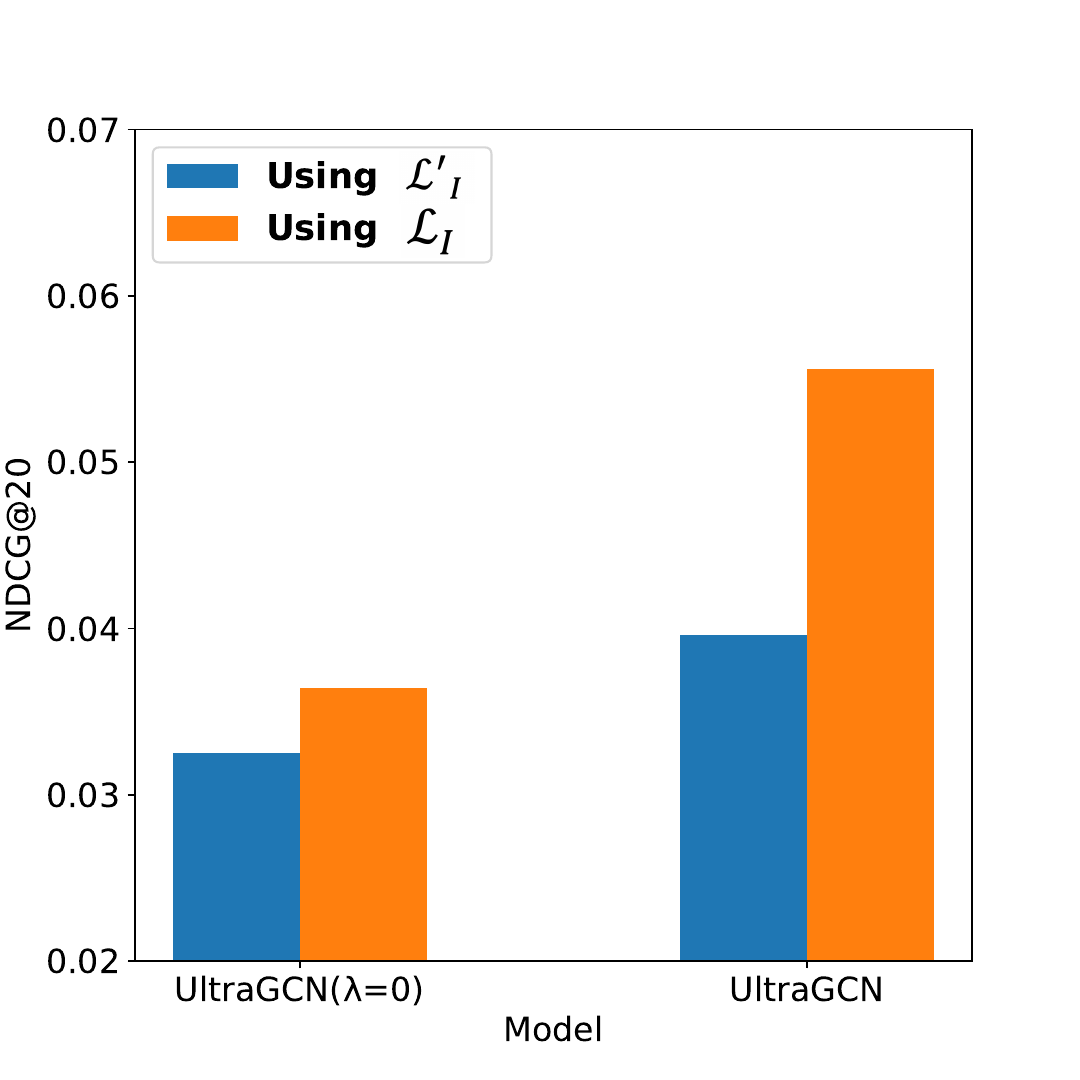}}
	\vspace{-2ex}
	\caption{Performance comparison of using $\mathcal{L'}_I$ and $\mathcal{L}_I$.}
	\label{fig_ablation_study_ii}
	\vspace{-2ex}
\end{figure}

\textbf{For opinion (ii)}, we change $\mathcal{L}_I$ to $\mathcal{L'}_{I}$:
\begin{equation}
	\mathcal{L'}_{I} = -\sum_{(u,i)\in N^+}\sum_{j \in S(i)} \omega_{i, j} \log(\sigma(e_{i}^{\top} e_{j}))
\end{equation}
which is instead to optimize between the target positive item and its most $K$ similar items. We compare the performance of UltraGCN using $\mathcal{L}_I$ and $\mathcal{L'}_I$ respectively with careful parameters tuning. Results are shown in Figure~\ref{fig_ablation_study_ii}. It is clear that no matter incorporating $\mathcal{L}_C$ or not, using $\mathcal{L}_I$ can achieves obvious better performance than using $\mathcal{L'}_I$, which proves that our designed strategy to learn on item-item graph is more effective. Furthermore, the performance gap between using $\mathcal{L}_I$ and using $\mathcal{L'}_I$ becomes large when incorporating $\mathcal{L}_C$, indicating that our strategy which makes the objective of UltraGCN unified can thus facilitate training and improve performance. 

\textbf{For opinion (iii)}, we derive the user-user constraint loss $\mathcal{L}_U$ with the similar method of Section~\ref{learning_on_ii} and combine it to the final objective. We carefully re-tune the parameters and show the comparison results of whether using $\mathcal{L}_U$ in Table~\ref{table::ablation_study_uu}. 
As can be seen, incorporating $\mathcal{L}_U$ to learn user-user relationships does not bring obvious benefits to UltraGCN. We attribute this phenomenon to the fact that the users' interests are broader than items' properties, and thus it is much harder to capture user-user relationships just from the user-user co-occurrence graph. Therefore, we do not introduce the modeling of user-user relationships into UltraGCN in this paper, and we will continue to study it in the future.


\subsection{Parameter Analysis}
We investigate the influence of the number of selected neighbors $K$ and the weights of the two constraint losses (i.e., $\lambda$ and $\gamma$) on the performance for a better understanding of UltraGCN.

\subsubsection{Impact of $K$}
\begin{figure}[!t]
	\centering  
	\subfigure[Amazon-Book]{
		\includegraphics[width=0.485\columnwidth]{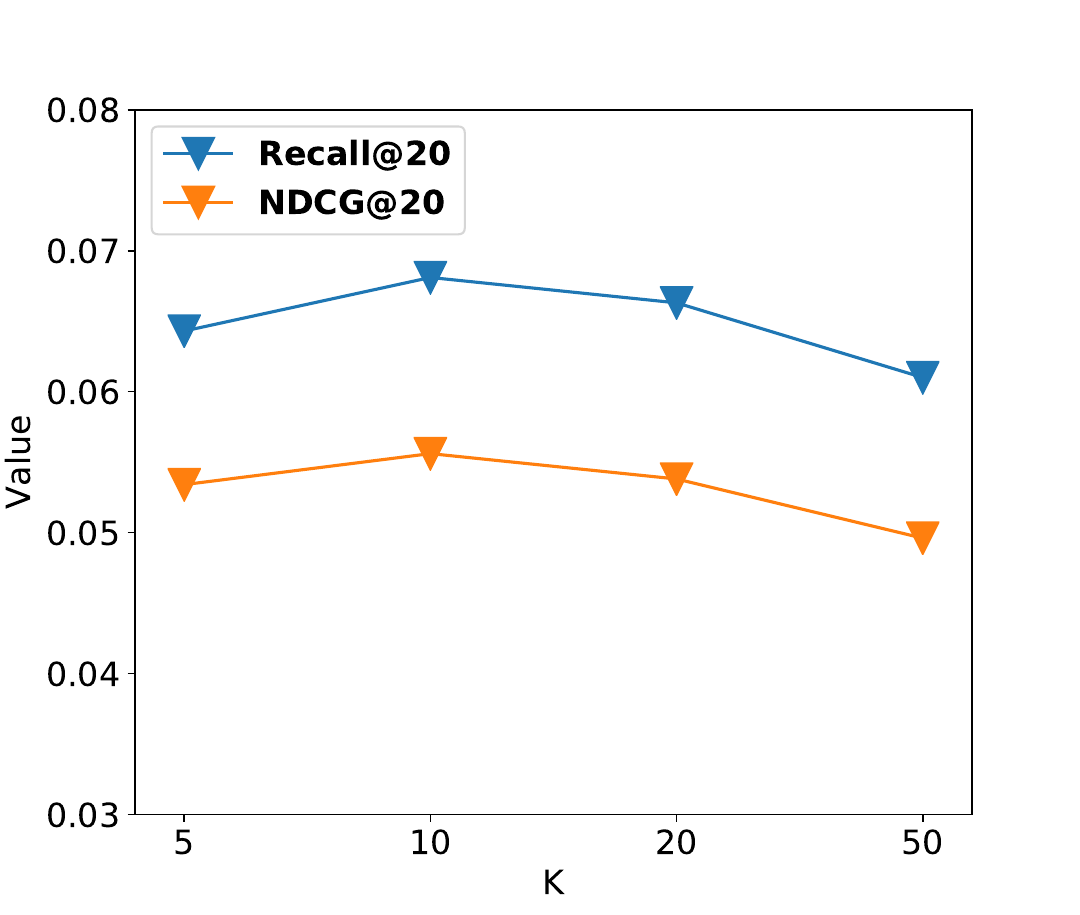}}
	\subfigure[Yelp2018]{
		\includegraphics[width=0.485\columnwidth]{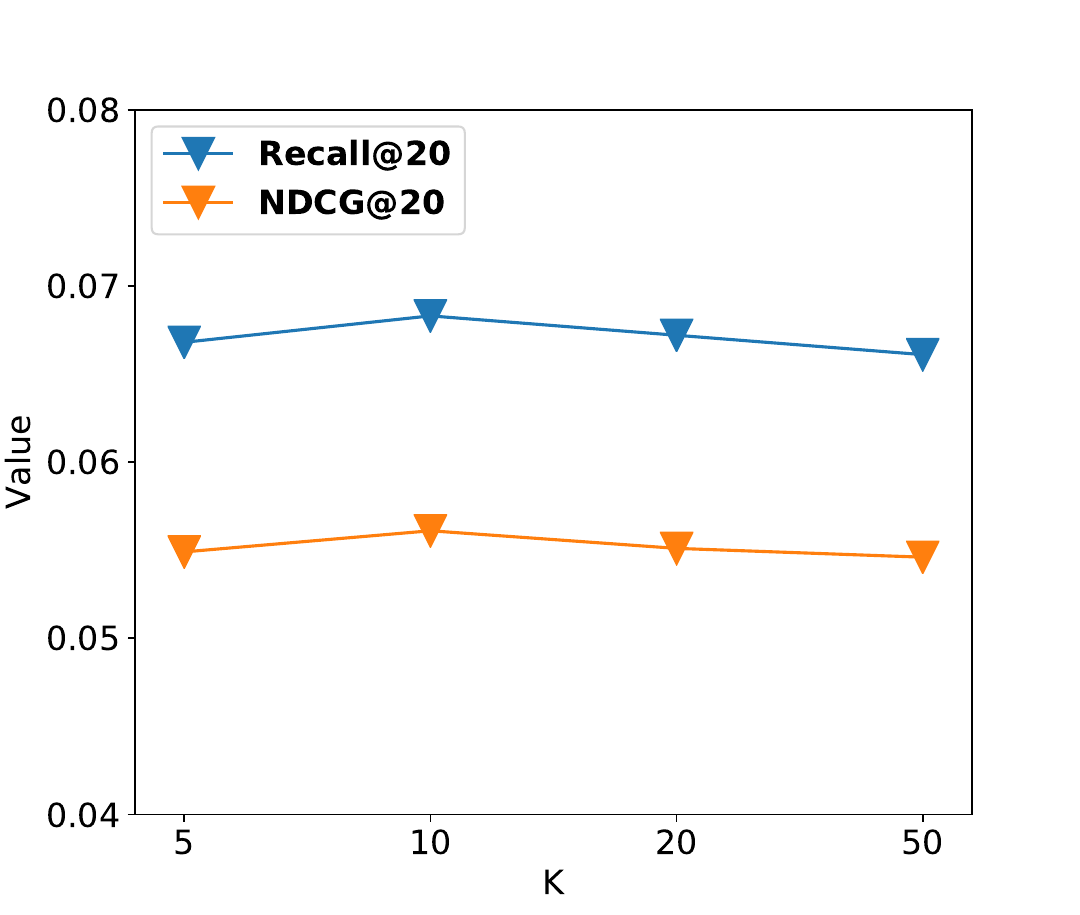}}
	\vspace{-2ex}
	\caption{Performance comparison of setting different $K$.}
	\vspace{-2ex}
	\label{param_k}
\end{figure}

We test the performance of UltraGCN with different $K$ in [5, 10, 20, 50] on Amazon-Book and Yelp2018. Figure~\ref{param_k} shows the experimental results. We can find that when $K$ increases from 5 to 50, the performance shows a trend of increasing first and then decreasing. This is because that when $K$ is 5, the item-item relationships are not sufficiently exploited. While when $K$ becomes large, there may introduce some less similar or less confident item-item relationships into the learning process that affect model performance. Such phenomenon also confirms that conventional GCN-based CF models inevitably take into account too many low-confidence relationships, thus hurting performance.  

\subsubsection{Impact of $\lambda$ and $\gamma$}
\begin{figure}[!t]
	\centering  
	\subfigure[Impact of $\lambda$]{
		\includegraphics[width=0.485\columnwidth]{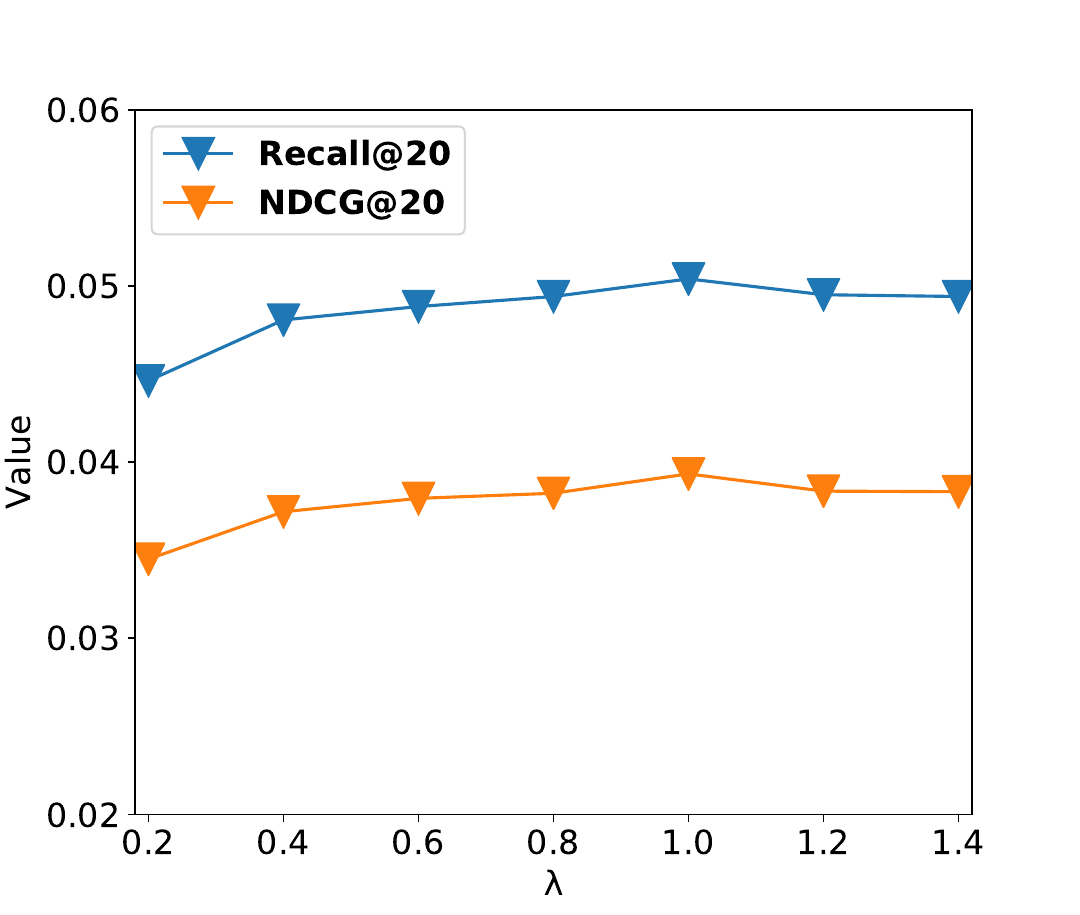}}
	\subfigure[Impact of $\gamma$]{
		\includegraphics[width=0.485\columnwidth]{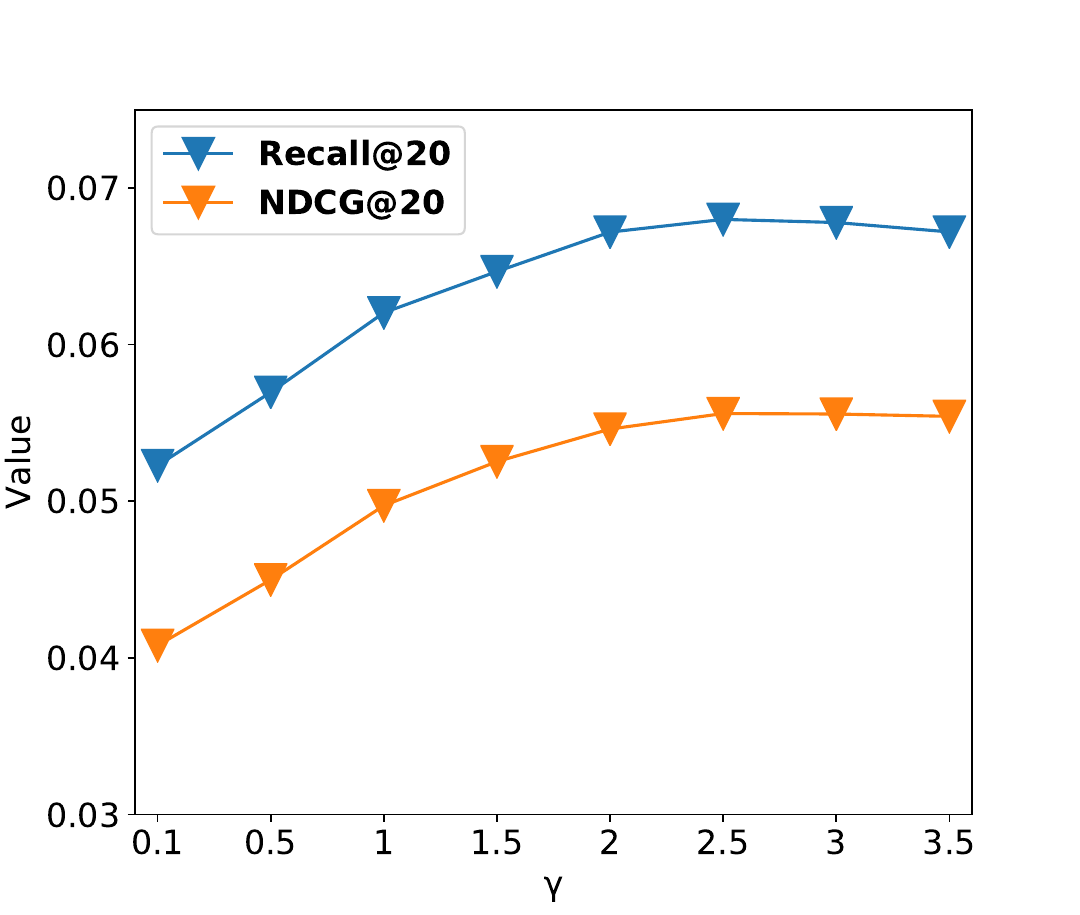}}
	\vspace{-2ex}	
	\caption{Performance comparison with different $\lambda$ and $\gamma$.}
	\label{param_lambda_gamma}
\end{figure}

We first set $\lambda=0$ and show the performance of different $\lambda$ from 0.2 to 1.4 (0.2 as the interval). Then we test with different $\gamma$ in [0.1, 0.5, 1, 1.5, 2, 2.5, 3, 3.5] based on the best $\lambda$. Experiments are conducted on Amazon-Book, and we show results in Figure~\ref{param_lambda_gamma}.
For $\lambda$, we find that the small value limits the exertion of the user-item constraint loss, and a value of 1 or so would be suitable for $\lambda$.
For the impact of $\gamma$, its trend is similar to $\lambda$ but is more significant, and 2.5 is a suitable choice for $\gamma$. In general, our investigations for $\lambda$ and $\gamma$ show that these two parameters are important to UltraGCN, which can flexibly adjust the learning weights for different relationships and should be carefully set.

%% file: sections/5_relatedwork.tex
\section{Related Work}
In this section, we briefly review some representative GNN-based methods and their efforts for model simplification toward recommendation tasks.

 With the development and success of GNN in various machine learning areas, there appears a lot of excellent work in recommendation community since the interaction of users and items could be naturally formed to a user-item bipartite graph.
 Rianne van den Berg et al.~\cite{GC-MC} propose graph convolutional matrix completion (GC-MC), a graph-based auto-encoder framework for explicit matrix completion. The encoder of GC-MC aggregates the information from neighbors based on the types of ratings, and then combine it to the new embeddings of the next layer. It is the first work using graph convolutional neural networks for recommendation. 
 Ying et al.~\cite{PinSage} first applys GCN on web-scale recommender systems and propose an efficient GCN-based method named Pinsage, which combines efficient random walks and graph convolutions
 to generate embeddings of items that incorporate both graph structure as well as item feature information.
 Then, Wang et al.~\cite{NGCF} design NGCF which is a new graph-based framework for collaborative filtering. NGCF has a crafted interaction encoder to capture the collaborative signals among users and items.  Although NGCF achieves good performance compared with previous non-GNN based methods, its heavy designs limit its efficiency and full exertion of GCN. To model the diversity of user intents on items, Wang et al.~\cite{DGCF} devise Disentangled Graph Collaborative Filtering (DGCF), which considers user-item relationships at the finer granularity of user intents and generates disentangled user and item representations to get better recommendation performance. 
 
Although GNN-based recommendation models have achieved impressive performance, their efficiencies are still unsatisfactory when facing large-scale recommendation scenarios. How to improve the efficiency of GNNs and reserve their high performance for recommendation becomes a hot research problem. Recently,  Dai et al.~\cite{fixed_point_GNN_dai} and Gu et al.~\cite{implicitGNN} extend fixed-point theory on GNN for better representation learning. Liu et al.~\cite{UCMF} propose UCMF that simplifies GCN for the node classification task.
Wu et al.~\cite{SGCN} find the non-necessity of nonlinear activation and feature transformation in GCN, proposing a simplified GCN (SGCN) model by removing these two parts. Inspired by SGC, He et al.~\cite{LightGCN} devise LightGCN for recommendation by removing nonlinear activation and feature transformation too. However, its efficiency is still limited by the time-consuming message passing. Qiu et al.~\cite{unify_network_embedding} demonstrate that many network embedding algorithms with negative sampling can be unified into the MF framework which may be efficient, however, their performances still have a gap between that of GCNs.
We are inspired by these instructive studies, and propose UltraGCN for both efficient and effective recommendation.

%% file: sections/6_conclusion.tex
\vspace{-1.5ex}
\section{Conclusion}
In this work, we propose an ultra-simplified formulation of GCN, dubbed UltraGCN. UltraGCN skips explicit message passing and directly approximate the limit of infinite message passing layers. 
Extensive experimental results demonstrate that UltraGCN achieves impressive improvements over the state-of-the-art CF models in terms of both accuracy and efficiency.

\section{Acknowledgement}
This work was supported in part by the National Natural Science Foundation of China (61972219), the Research and Development Program of Shenzhen (JCYJ20190813174403598, SGDX20190918101201696), the National Key Research and Development Program of China (2018YFB1800601), and the Overseas Research Cooperation Fund of Tsinghua Shenzhen International Graduate School
(HW2021013).

%% file: sections/7_appendix.tex
\section{Appendix}

To further demonstrate the effectiveness of UltraGCN, we additionally provide the results compared to some more recent state-of-the-art CF models, including NBPO~\cite{NBPO}, BGCF~\cite{BGCF}, SCF~\cite{SpectralCF}, LCFN~\cite{LCFN}, and SGL-ED~\cite{SGL-ED}. For simplicity and fairness of comparison, we use the same dataset and evaluation protocol provided by each paper. We also duplicate the results reported in their papers to keep consistency. The results in Table~\ref{addtional_res1} again validate the effectiveness of UltraGCN, which outperforms the most recent CF models by a large margin.

\vspace{-1.5ex}

\begin{table}[H]
\centering
\caption{Performance comparison with some more models, including SCF, LCFN, NBPO,  BGCF, and SGL-ED.} \label{addtional_res1}

{\setlength{\tabcolsep}{7pt}
\begin{tabular}{ccc}
\hline
\multicolumn{3}{c}{Movielens-1M}      \\ \hline
Model     & F1@20           & NDCG@20         \\ \hline
NGCF     & 0.1582          & 0.2511          \\
SCF      & 0.1600          & 0.2560          \\
LCFN      & \underline{0.1625}          & \underline{0.2603}          \\ \hline
UltraGCN  & \textbf{0.2004} & \textbf{0.2642} \\
Improv. & 23.3\%           & 1.5\%           \\ \hline
\end{tabular}
}

\begin{tabular}{c}
~\vspace{0ex}
\end{tabular}

{\setlength{\tabcolsep}{7pt}
\begin{tabular}{ccc}
\hline
\multicolumn{3}{c}{Amazon-Electronics}      \\ \hline
Model    & F1@20           & NDCG@20         \\ \hline
MF-BPR   & 0.0275          & 0.0680          \\
ENMF     & \underline{0.0314}          & \underline{0.0823}          \\
NBPO     & 0.0313          & 0.0810          \\ \hline
UltraGCN & \textbf{0.0330} & \textbf{0.0829} \\
Improv.  & 5.1\%           & 0.7\%           \\ \hline
\end{tabular}
}

\begin{tabular}{c}
~\vspace{0ex}
\end{tabular}

\begin{tabular}{ccc}
\hline
\multicolumn{3}{c}{Amazon-CDs}               \\ \hline
Model    & Recall@20       & NDCG@20         \\ \hline
NGCF     & 0.1258          & 0.0792          \\
NIA-GCN  & 0.1487          & 0.0932          \\
BGCF     & \underline{0.1506}          & \underline{0.0948}          \\ \hline
UltraGCN & \textbf{0.1622} & \textbf{0.1043} \\
Improv.  & 7.7\%           & 10.0\%          \\ \hline
\end{tabular}

\begin{tabular}{c}
~\vspace{0ex}
\end{tabular}

\begin{tabular}{ccc}
\hline
\multicolumn{3}{c}{Amazon-Books}               \\ \hline
Model    & Recall@20       & NDCG@20         \\ \hline
NGCF     & 0.0344          & 0.0263          \\
LightGCN  & 0.0411          & 0.0315          \\
SGL-ED     & \underline{0.0478}         & \underline{0.0379}          \\ \hline
UltraGCN & \textbf{0.0681} & \textbf{0.0556} \\
Improv.  & 42.5\%           & 46.7\%          \\ \hline
\end{tabular}

\end{table}


